\documentclass[aps,prb,twocolumn]{revtex4-1}

\usepackage[mathscr]{euscript}
\usepackage{enumitem}
\usepackage{amsmath,bm}
 \usepackage{graphicx}
\usepackage{hyperref}
\hypersetup{
  colorlinks = true,
  citecolor = blue,
  urlcolor = blue
}
\newcommand{{\bfx}}{\mbox{\boldmath$x$\unboldmath}}
\newcommand{{\bfk}}{\mbox{\boldmath$k$\unboldmath}}
\newcommand{{\bfv}}{\mbox{\boldmath$v$\unboldmath}}
\newcommand{{\bfq}}{\mbox{\boldmath$q$\unboldmath}}
\newcommand{{\bfp}}{\mbox{\boldmath$p$\unboldmath}}
\newcommand{{\bfa}}{\mbox{\boldmath$a$\unboldmath}}
\newcommand{{\bfT}}{\mbox{\boldmath$T$\unboldmath}}
\newcommand{{\bfE}}{\mbox{\boldmath$E$\unboldmath}}
\newcommand{{\bfm}}{\mbox{\boldmath$m$\unboldmath}}
\newcommand{{\bfe}}{\mbox{\boldmath$e$\unboldmath}}
\newcommand{{\bfc}}{\mbox{\boldmath$c$\unboldmath}}
\newcommand{{\bfg}}{\mbox{\boldmath$g$\unboldmath}}
\newcommand{{\bff}}{\mbox{\boldmath$f$\unboldmath}}
\newcommand{{\bfF}}{\mbox{\boldmath$F$\unboldmath}}
\newcommand{{\bfA}}{\mbox{\boldmath$A$\unboldmath}}
\newcommand{{\bfmu}}{\mbox{\boldmath$\mu$\unboldmath}}
\newcommand{{\bfOmega}}{\mbox{\boldmath$\Omega$\unboldmath}}

\newcommand{{\bfbeta}}{\mbox{\boldmath$\beta$\unboldmath}}
\newcommand{{\bfphi}}{\mbox{\boldmath$\phi$\unboldmath}}
\newcommand{{\bfrho}}{\mbox{\boldmath$\rho$\unboldmath}}
\newcommand{{\bfcalL}}{\mbox{\boldmath$\cal L$\unboldmath}}
\newcommand{{\bfcalK}}{\mbox{\boldmath$\cal K$\unboldmath}}
\newcommand{\barlambda}{{\mkern0.75mu\mathchar '26\mkern -9.75mu\lambda}}
\newcommand{\gradv}{\boldsymbol{\nabla}}

\begin{document}


\title{The classical counterpart of the Aharonov-Bohm phase}

\author{Ricardo Heras}
\email[]{ricardo.heras.13@ucl.ac.uk}

\affiliation{Department of Physics and Astronomy, \\University College London, London WC1E 6BT, UK}

\begin{abstract}
\noindent The existence of the Aharonov-Bohm phase shows that the magnetic field may produce nonlocal effects in quantum mechanics. It is generally believed that such a nonlocal behavior of the magnetic field is not possible in classical physics and that this represents a clear difference between classical and quantum mechanics. Contrary to these beliefs, we argue that the classical counterpart of the Aharonov-Bohm phase, which is identified here with the electromagnetic angular momentum of the configuration formed by an electric charge moving around an infinitely-long solenoid enclosing a uniform magnetic flux, shows that the magnetic field may produce nonlocal effects in classical mechanics. We discuss this momentum in detail by putting special emphasis on its topological and nonlocal features. The momentum is topological because it depends on the number of windings the electric charge carries out around the solenoid and is nonlocal because the magnetic flux has no local consequences at any point on the charge trajectory. The topological feature allows us to introduce the concept of accumulated electromagnetic angular momentum and the nonlocal feature allows us to speak of a dynamical nonlocality attributable to the equations of classical electrodynamics.

\end{abstract}

\maketitle

\section{Introduction}\label{I}
\noindent In quantum mechanics, the wave function of an electric charge encircling an infinitely-long solenoid enclosing a uniform magnetic flux acquires a quantum phase, the famous Aharonov-Bohm (AB) phase.\cite{1} The point to remark is that the charge moves in a non-simply connected region, where there is no magnetic field --and therefore there is no Lorentz force acting on the charge-- but there is a vector potential. Two of the main features of the AB phase are its topology and nonlocality.\cite{2} The AB  phase is topological because it depends on the number of windings the charge carries out around the solenoid and is independent of the shape of the charge trajectory. The AB  phase is nonlocal because there are no local consequences at any point of the charge trajectory due to the magnetic flux. This nonlocal feature of the AB phase is identified with a dynamical quantum nonlocality\cite{3,4} which has been attributed to the quantum equations of motion.\cite{3,4,5}

Considering that the AB phase is distinctively quantum mechanical, the identification of its classical counterpart is a challenging task. In this regard, it is pertinent to distinguish the idea of a classical counterpart of the AB phase from the idea of classical analogues to the AB effect.
The classical counterpart must share the same electromagnetic configuration used in the determination of the AB phase and exhibit some of the properties of this phase like its topology and nonlocality. On the other hand, several classical analogues of the AB effect have been proposed\cite{2,6,7,8,9,10} but none of them involve the same electromagnetic configuration of the AB phase.

The natural candidate to represent the classical counterpart of the AB phase is the electromagnetic angular momentum of the configuration formed by an electric charge $q$ encircling an infinitely-long solenoid with the uniform magnetic flux $\Phi$:
\begin{align}
\mathbf{L}_\texttt{AB}= \frac{q\Phi\hat{\mathbf{z}}}{2\pi c}\,.
\end{align}
 Here we are using cylindrical coordinates and Gaussian units. The remarkable point is that the magnitude of this momentum: $L_\texttt{AB}=q\Phi/(2\pi c) $ is related to the AB phase: $\delta_\texttt{AB}=q\Phi/(\hbar c)$ through the linear relation:
\begin{align}
\delta_\texttt{AB}=\frac{2\pi}{\hbar} L_\texttt{AB},
\end{align}
which naturally connects a quantum prediction with a classical prediction and strongly suggests that the classical  momentum $L_\texttt{AB}$ could share the same topological and nonlocal features of the quantum phase $\delta_\texttt{AB}.$

The momentum given in Eq.~(1) has been derived using two different approaches. On one hand, Peshkin\cite{11,12} and more recently Tiwari\cite{13} derived this momentum using a Lagrangian treatment. On the other hand, Wakamatsu et al.\cite{14} recently obtained  this momentum using the time-independent formula \cite{15}
\begin{align}
\textbf{L}_{\texttt{M}}= \frac{1}{c}\int_{V}\mathbf{x} \times (\rho_\texttt{e} \mathbf{A}) \,d^3x,
\end{align}
where $\rho_\texttt{e}$ is a localised charge density and \textbf{A} is a vector potential in the Coulomb gauge ($\gradv \cdot \textbf{A}= 0$). We will call Eq.~(3) the Maxwell formula.

Nevertheless, none of the previously mentioned authors discussed the topological and nonlocal features of the momentum  $L_\texttt{AB}$ nor examined the idea that it can be considered to be the classical counterpart of the AB phase $\delta_\texttt{AB}.$ Furthermore, Wakamatsu et al.\cite{14} obtained the relation given in Eq.~(2) but without providing any further discussion on the fact that it naturally connects classical and quantum predictions. The only antecedent for this relation we are aware of, is the generic formula $\delta_\texttt{AB}=(1/\hbar)\oint \hat{\mathbf{z}}\cdot \mathbf{L}_\texttt{em}d\phi$ suggested several years ago by Maeda and Shizuya.\cite{16} In the last formula $\mathbf{L}_\texttt{em}$ represents the electromagnetic angular momentum of a configuration formed by a charge and a magnetic moment.

Despite its evident conceptual importance, the momentum $\mathbf{L}_\texttt{AB}$ in Eq.~(1) has not been closely examined. Some fundamental questions that deserve to be clarified are the following: (i)  Can the standard formula  for the electromagnetic angular momentum, which we will call the Poynting formula,
\begin{align}
\textbf{L}_{\texttt{P}}= \frac{1}{4\pi c}\int_{V}\mathbf{x} \times (\mathbf{E}\times \mathbf{B}) \,d^3x,
\end{align}
be used to obtain $\mathbf{L}_\texttt{AB}$? (ii) How are explicitly related the Maxwell and Poynting formulas given in Eqs.~(3) and (4)? (iii) Is the Maxwell formula in Eq.~(3) gauge invariant? (iv) Does the momentum $\mathbf{L}_\texttt{AB}$ share the topological and nonlocal features of the AB phase $\delta_\texttt{AB}$?
(v) Can the momentum $\mathbf{L}_\texttt{AB}$ be considered the classical counterpart of the AB phase $\delta_\texttt{AB}$? (vi) Is it crucial for the momentum $\mathbf{L}_\texttt{AB}$ that the AB configuration is defined in a non-simply connected region? (vii) Could we conclude that the momentum $\mathbf{L}_\texttt{AB}$ is a nonlocal manifestation of the magnetic field in classical physics?

In this paper we discuss in detail the momentum $\mathbf{L}_\texttt{AB}$ by putting special emphasis on its topological and nonlocal features. We introduce a new relation that connects the Poynting and Maxwell formulas for the electromagnetic angular momentum.  We stress the essential role that plays the non-simply connected region where this momentum is defined. We argue that this momentum can be considered to be the classical counterpart of the AB phase, suggest how  it can be physically manifested and emphasise that its existence shows that the magnetic field can produce nonlocal effects in classical physics.

The plan of the paper is as follows. In Sec.~\ref{II} we describe the electromagnetic AB configuration. In Sec.~\ref{III} we decompose the Poynting formula into a volume part (the Maxwell formula) and a surface part. We show that in the case of the electromagnetic AB configuration, the surface part plays the role of a return momentum at the infinite surface. In Sec.~\ref{IV} we derive the momentum  $\mathbf{L}_\texttt{AB}$ and discuss its topological character. We introduce the idea of an accumulated electromagnetic angular momentum and discuss its independence of the dynamics of the charged particle. We also discuss the nonlocal feature of the momentum $\mathbf{L}_\texttt{AB}$ and the idea of a dynamical classical nonlocality attributable to equations of classical electrodynamics. In Sec.~\ref{V} we present a discussion between Alice and Bob about the nonlocality of the momentum  $\mathbf{L}_\texttt{AB}$.
In Sec.~\ref{VI} we present a general proof that the Maxwell formula is gauge invariant when applied to two dimensional configurations like the AB configuration.
In Sec.~\ref{VII} we discuss the measurability of the momentum $\mathbf{L}_\texttt{AB}$. In Sec.~\ref{VIII} we discuss the relation between the momentum  $\mathbf{L}_\texttt{AB}$ and the AB phase $\delta_\texttt{AB}$ which enlightens the idea that the former may be considered the classical counterpart of the latter. As a conclusion, we answer in Sec.~\ref{IX} the questions (i)-(vii) raised in the introduction. In Appendix A we derive the AB potentials (inside and outside the solenoid). In Appendix B we verify an identity used in the decomposition of the Poynting formula and in Appendix C we derive the return momentum  $-\mathbf{L}_\texttt{AB}$ of the AB configuration.

\noindent \section{The electromagnetic AB configuration}\label{II}
\noindent The electromagnetic AB configuration consists of an electric charge $q$ moving around
an infinitely-long solenoid laying along the $z$-axis, of circular cross-section with radius $R$, enclosing the uniform magnetic flux $\Phi=\pi R^2 B$, where $B$ is the magnitude of the magnetic field.  The current density of this configuration in cylindrical coordinates reads\cite{17} $\textbf{J}=c\Phi\delta(\rho-R)\hat{\bfphi}/(4\pi^2 R^2),$ where $\delta$ is the Dirac delta function. Notice that this current is a magnetisation current: $\textbf{J}=c\gradv\times \textbf{M}$, where the magnetisation vector is given by $\textbf{M}=\Phi\Theta(R-\rho)\hat{\textbf{z}}/(4\pi^2 R^2)$ with $\Theta$ being the Heaviside step function with values $\Theta=1$ if $R>\rho$ and $\Theta =0$ if $R<\rho$. The corresponding Maxwell's equations are given by
\begin{align}
\gradv\cdot \textbf{B}=0, \quad \gradv\times \textbf{B} = \frac{\Phi\delta(\rho-R)\hat{\bfphi}}{\pi R^2}\, .
\end{align}
We can directly verify that the corresponding solution is
\begin{align}
\textbf{B}= \frac{\Phi\Theta(R-\rho)\hat{\textbf{z}}}{\pi R^2}.
\end{align}
From Eq.~(6) we observe that the magnetic field is confined within the solenoid and it is connected with the magnetisation vector through the relation $\textbf{B}=4\pi\textbf{M}.$ Inserting $\textbf{B}=\gradv \times  \textbf{A},$ using $\gradv^2 \textbf{A}\equiv \gradv(\gradv \cdot  \textbf{A})-\gradv \times (\gradv \times  \textbf{A}),$ and adopting the Coulomb gauge $\gradv \cdot  \textbf{A}=0$ the second equation of Eqs.~(5) implies the Poisson equation
\begin{align}
\gradv^2 \textbf{A} = - \frac{\Phi\delta(\rho-R)\hat{\bfphi} }{\pi R^2}.
\end{align}
The solution of this Poisson equation is given by\cite{18}
\begin{align}
\textbf{A}=\bigg(\frac{\Phi}{2\pi\rho}\Theta(\rho-R)+\frac{\Phi\rho}{2\pi R^2}\Theta(R-\rho)\bigg)\hat{\bfphi}.
\end{align}
An explicit derivation of Eq.~(8) is presented in Appendix A. As may be seen, the potential in Eq.~(8) is not defined at $\rho=R$, that is, at the solenoid surface. However, in Appendix A we regularise Eq.~(8) and show that $\textbf{A}(R)= \Phi\hat{\bfphi}/(2\pi R)$ which indicates that the potential is continuous at $\rho=R.$ Outside the solenoid $(\rho>R)$ the magnetic field vanishes and the potential $\textbf{A}$ reads
\begin{align}
\mathbf{A} = \frac{\Phi}{2 \pi}\frac{\hat{\bfphi}}{\rho}.
\end{align}
This potential can be expressed as $\mathbf{A}= \gradv \chi,$ where $\chi=(\Phi/2 \pi)\phi$ with $\phi$ being the azimuthal coordinate. The scalar function $\chi$ satisfies $\gradv^2\chi=0.$ We also note that $\chi$ is a multi-valued function [that is, $\chi(\phi)\not=\chi(\phi+2\pi)]$ and that its gradient $\gradv \chi$ is a singular function. The region outside the solenoid is a force-free region since the Lorentz force $\mathbf{F}\! = \!q\, \textbf{\emph{v}}\times \mathbf{B}/c$, where $\emph{\textbf{v}}$ is the charge velocity, vanishes because the magnetic field is zero in this region.

\section{Decomposing the momentum {\large $\mathbf{\texttt{L}}_\texttt{AB}$}}\label{III}
\noindent When applying the Poynting formula given in Eq.~(4) to calculate the momentum $\mathbf{L}_\texttt{AB}$ outside the solenoid we find that $\mathbf{L}_\texttt{AB}=0$ because the field $\textbf{B}$ is zero in this region. Nevertheless, when applying the Maxwell formula given in Eq.~(3) outside the solenoid we find the non-vanishing momentum $\textbf{L}_{\texttt{AB}}=q\Phi\hat{\textbf{z}}/(2\pi c).$ Furthermore, a Lagrangian treatment supports the result obtained by the Maxwell formula. In fact, the Lagrangian ${\cal L}=m \emph{\textbf{v}}^2/2 + q\emph{\textbf{v}}\cdot \textbf{A}/c$, with $\mathbf{A} = \Phi\hat{\bfphi}/(2 \pi\rho)$, leads to the canonical angular momentum of the electromagnetic AB configuration:\cite{9} $\textbf{L}_{\texttt{can}}= \mathbf{x} \times \textbf{p}_{\texttt{can}}= \textbf{L}_{\texttt{mech}}+ \textbf{L}_{\texttt{AB}}$, where $\textbf{L}_{\texttt{mech}}=\textbf{x}\times (m\emph{\textbf{v}})$ is the mechanical angular momentum and $\textbf{L}_{\texttt{AB}}=q\Phi\hat{\textbf{z}}/(2\pi c)$ is the corresponding electromagnetic angular momentum. Furthermore, the canonical momentum $\textbf{L}_{\texttt{can}}$ coincides with the total angular momentum, which is shown to be conserved.\cite{9} The question arises: Is the Poynting formula inappropriate to calculate the momentum $\textbf{L}_{\texttt{AB}}$?

Let us decompose the Poynting formula to answer this question. In the Appendix B  we verify the identity
\begin{align}
\nonumber [\mathbf{x} \times(\mathbf{E}\times \mathbf{B})]^\texttt{s}=&[4\pi\,\mathbf{x} \times (\rho_\texttt{e} \mathbf{A})]^\texttt{s} \\+&
\partial^\texttt{k}[\varepsilon^{\texttt{sqi}}x_\texttt{q}(\delta_\texttt{ik}E_\texttt{m}A^\texttt{m}\!-\!E_\texttt{k}A_\texttt{i}\!-\!E_\texttt{i}A_\texttt{k})],
\end{align}
which holds for stationary systems involving Coulomb-gauge potentials. In this identity $\rho_\texttt{e}$ is a  localised charge density, $\partial^\texttt{k}= (\gradv)^\texttt{k}, \varepsilon^{\texttt{sqi}}$ is the Levi-Civita tensor, $\delta_\texttt{ik}$ is the Kronecker delta and summation on repeated indices is understood.  The volume integration of Eq.~(10) gives two volume integrals on the right-hand side. The last of these integrals can be transformed into a surface integral. The final result is the decomposition
\begin{align}
\textbf{L}_\texttt{P}=\textbf{L}_\texttt{M}+\textbf{L}_\texttt{S},
\end{align}
where
\begin{align}
\textbf{L}_\texttt{P}&=\frac{1}{4\pi c}\int_{V}\mathbf{x} \times (\mathbf{E}\times \mathbf{B})\,d^3x,\\
\textbf{L}_\texttt{M}&=\frac{1}{c}\int_{V}\mathbf{x} \times (\rho_\texttt{e} \mathbf{A})\,d^3x, \\
\textbf{L}_\texttt{S}&=\frac{1}{4\pi c}\!\oint_{S} \mathbf{x} \!\times\!\big[\hat{\textbf{n}}(\textbf{E}\!\cdot\! \textbf{A})\!-\!\textbf{A}(\hat{\textbf{n}}\!\cdot\!\textbf{E})\!-\!\textbf{E}(\hat{\textbf{n}}\!\cdot\! \textbf{A})\big]d{S}.
\end{align}
Here $S$ denotes the surface of the volume $V$. As far we  are aware, the formula given in Eq.~(11) is new. According to this  formula the momentum $\textbf{L}_\texttt{P}$ can be decomposed into a part $\textbf{L}_\texttt{M}$ defined in the volume ${V}$ which identifies with the Maxwell formula plus another part $\textbf{L}_{\texttt{S}}$ defined on the surface $S$ of the volume ${V}$.

Now we can apply the decomposition in Eq.~(11) to the electromagnetic angular momentum of the AB configuration. In this case we have $\textbf{L}_\texttt{P}=0$ because the magnetic field $\textbf{B}$ is zero outside the solenoid. Therefore, $0=\textbf{L}_\texttt{M}+\textbf{L}_\texttt{S}$, which implies $\textbf{L}_\texttt{M}=-\textbf{L}_\texttt{S}.$ Accordingly, in the electromagnetic AB configuration there is the momentum $\textbf{L}_\texttt{M}$ in the infinite spatial region where the charge $q$ is moving around the solenoid and there is the momentum $\textbf{L}_\texttt{S}=-\textbf{L}_\texttt{M}$ on the surface of this region which lies at infinity. Put differently, when calculating the momentum $\textbf{L}_\texttt{M}$ in the electromagnetic AB configuration, we must consider its return momentum $\textbf{L}_\texttt{S}=-\textbf{L}_\texttt{M}$ to ensure the momentum $\textbf{L}_\texttt{P}$ is zero outside the solenoid. But there is no physical inconsistence because the return momentum  $\textbf{L}_\texttt{S}$ is defined in a different region (the infinite surface of the volumetric space outside the solenoid) to that where the charge $q$  is moving (the volumetric space outside the solenoid). The Poynting formula is appropriate to calculate the momentum of the electromagnetic AB configuration whenever this formula is decomposed in its volumetric and surface parts.

\section{Topological and nonlocal features of the momentum {\large $\mathbf{\texttt{L}}_\texttt{AB}$}}\label{IV}
\begin{figure}
  \centering
  \includegraphics[width= 148pt]{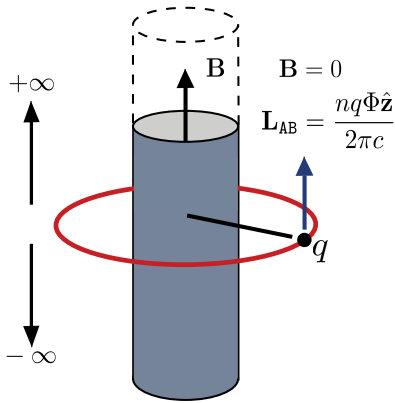}
  \caption{The system formed by a charge encircling an infinitely-long solenoid enclosing a uniform magnetic flux has an electromagnetic angular momentum which is topological, nonlocal, and gauge-invariant.}\label{Fig1}
\end{figure}

\noindent Let us apply the Maxwell formula expressed as
\begin{align}
\mathbf{L}_\texttt{M}= \frac{1}{c}\int_{{V}}\,\mathbf{x}' \times \big(\rho_{\texttt{e}}(\mathbf{x}')\mathbf{A(\mathbf{x}')}\big)\,d^3x',
\end{align}
to calculate the electromagnetic angular momentum $\mathbf{L}_\texttt{AB}$. Inserting the quantities $\rho_{\texttt{e}}(\mathbf{x}')= q \delta(\rho'-\rho)\delta(z')/(2 \pi \rho'),$ $\mathbf{A(\mathbf{x}')}= \Phi \hat{{\bfphi}}/(2 \pi \rho')$ and $\mathbf{x'}= \rho' \hat{\bfrho}$
 in Eq.~(15) and integrating outside the solenoid we obtain
\begin{align}
\nonumber \mathbf{L}_{\texttt{AB}}=& \frac{q \Phi \hat{\mathbf{z}}}{4 \pi^2 c}\int_{-\infty}^{+\infty}\!\delta(z')\, dz'\int_{R}^\infty\!\delta(\rho'-\rho)\,d\rho' \oint\, d\phi'\\
=& \frac{q \Phi \Theta(\rho-R)\hat{\mathbf{z}}}{4 \pi^2 c} \oint d\phi'=\frac{q \Phi\hat{\mathbf{z}}}{4 \pi^2 c} \oint d\phi',
\end{align}
where $\Theta(\rho-R)=1$ because the charge is moving outside the solenoid, that is, $\rho>R$. In calculating the remaining integral $\oint\! d\phi'$  we must consider the fact that the charge is moving around the solenoid, that is, the charge can encircle the solenoid several times. This observation tacitly introduces topology. As the charge moves around the solenoid (see Fig.~\ref{Fig1}), the azimuthal angle $\phi'$ must vary continuously in time; thus we allow it to take on all values, and do not restrict it to lie in the interval $[0,2\pi],$ that is, $\oint_{C} d\phi'=\int_{t_\texttt{0}}^{t_\texttt{1}}(d\phi'/dt)dt=\phi'(t_\texttt{1})\!-\!\phi'(t_\texttt{0}),$ which expresses the net change in the angle $\phi'$ as the charge $q$ moves in the curve $C$ around the solenoid. Since the curve $C$ is closed, this change must be an integer multiple of $2\pi$, that is, $ \phi'(t_\texttt{1})-\phi'(t_\texttt{0})=2\pi n$, where $n$ is the winding number of the charge path $C$. It follows that the accumulated integration over $\phi'$ is $\oint\!d\phi'=2\pi n$, and therefore Eq.~(16) gives the {\it accumulated} electromagnetic angular momentum
\begin{align}
\mathbf{L}_\texttt{AB}=\frac{nq\Phi\hat{\mathbf{z}}}{2\pi c}.
\end{align}
We note that the accumulation of an electromagnetic angular momentum is a new concept in classical electrodynamics. Equation (17) can be expressed equivalently as $L_\texttt{AB}=nq\Phi/(2\pi c),$ on the understanding that it denotes the $z$-component of this momentum. The explicit form of the corresponding return momentum lying on the infinite surface is given by $\mathbf{L}^\texttt{S}_\texttt{AB}=- nq\Phi\hat{\mathbf{z}}/(2\pi c).$ This result is derived in Appendix C. Evidently, $\mathbf{L}_\texttt{AB}+ \mathbf{L}^\texttt{S}_\texttt{AB}=0$ as predicted by the Poynting formula outside the solenoid.

Under the previous considerations, we can emphasize that the momentum $L_\texttt{AB}$ is topological because it depends on the number of windings the charge carries out around the solenoid and is independent of the shape of the charge trajectory. The latter property implies that $L_\texttt{AB}$
is also independent of the dynamics of the charged particle (the only connection with dynamics deals with its dependence with respect to the winding number $n$).

To verify the independence of $\mathbf{L}_{\texttt{AB}}$ with regard to  the particle dynamics, consider the AB configuration where
the particle of mass $m$ and charge $q$ moves around the solenoid with the velocity $\textbf{\textit{v}}=(d\rho/dt)\hat{\bfrho}+\rho(d\phi/dt)\hat{\bfphi}$ (for simplicity consider a single turn: $n=1$). When this velocity and  the potential $\mathbf{A} = \Phi\hat{\bfphi}/(2 \pi\rho)$ are used in the canonical momentum of this dynamical configuration: $\textbf{P}_\texttt{can}=m\textbf{\textit{v}}+q \textbf{A}/c$, it takes the form
\begin{align}
\mathbf{P}_\texttt{can} =m\bigg(\frac{d\rho}{dt}\hat{\bfrho}+\rho\frac{d\phi}{dt}\hat{\bfphi}\bigg)
+\frac{q\Phi}{2 \pi c \rho}\hat{\bfphi}.
\end{align}
The canonical angular momentum is $\textbf{L}_\texttt{can}= \textbf{x}\times\mathbf{P}_\texttt{can}$.  Using Eq.~(18) and
$\textbf{x}=\rho\hat{\bfrho}$ we obtain
\begin{align}
\textbf{L}_\texttt{can}=m\rho^2\frac{d\phi}{dt}\hat{\textbf{z}} + \frac{q \Phi}{2 \pi c }\hat{\mathbf{z}}.
\end{align}
The first term on the right-hand side of Eq.~(19) is the mechanical angular momentum $\textbf{L}_\texttt{mech}=m\rho^2(d\phi/dt)\hat{\textbf{z}}$ and the second term is the electromagnetic angular momentum of the AB configuration $\textbf{L}_\texttt{AB}=q\Phi\hat{\textbf{z}}/(2\pi c)$. It is now evident that the momentum  $\textbf{L}_\texttt{AB}$ is independent of the motion of the charged particle.
\begin{figure}
  \centering
  \includegraphics[width= 150pt]{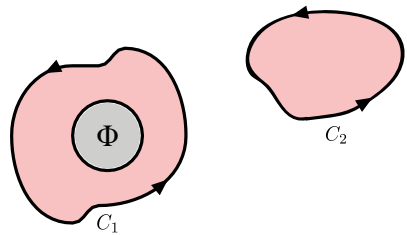}
  \caption{When the circulation of the potential \textbf{A} along the path $C_1$ encloses the solenoid we have a configuration defined in a non-simply connected region in which the electromagnetic angular momentum is non-zero. When the circulation of \textbf{A} along the path $C_2$ does not surround the solenoid we have a configuration defined in a simply-connected region in which the electromagnetic angular momentum vanishes.}\label{Fig2}
\end{figure}

It is important to emphasize that the AB configuration is defined in a \emph{non-simply} connected region and that this becomes crucial for the existence of the momentum $\textbf{L}_{\texttt{AB}}$. For example, if the charge would not encircle the solenoid then the corresponding configuration would be in a \emph{simply}-connected region outside the solenoid. This is so because the line of singularity at $\rho=0$ (that is along the entire $z$-axis) is excluded in this region (see Fig.~\ref{Fig2}). In this case the momentum in Eq.~(16) would vanish. To see this, we can write the momentum in Eq.~(16) as
\begin{align}
\mathbf{L}_\texttt{AB}= \frac{1}{c}\int_{{V}}\mathbf{x}' \times \big(\rho_{\texttt{e}}\mathbf{A}\big)\,d^3x' = \frac{q \hat{\textbf{z}}}{2\pi c}\oint_{C}\textbf{A}\cdot d\textbf{l}'.
\end{align}
From a well-known result of calculus in non-simply connected domains\cite{19} we can establish the following result:
\begin{align}
\oint_{C}\textbf{A}\cdot d\textbf{l}' = \begin{cases} n\Phi,\,\text{if $C$ encloses the solenoid}\\ 0, \, \text{otherwise}\end{cases}
\end{align}
Here $\textbf{A}$ is the AB potential and $n$ is the winding number. From Eqs.~(20) and (21) we note that the momentum $\textbf{L}_{\texttt{AB}}$ vanishes when the charge does not surround the solenoid.

A second fundamental feature of the momentum $L_\texttt{AB}$ is its nonlocality. This momentum is nonlocal because there is no force acting on the charge originated by the magnetic flux and therefore this flux has no local consequences at any point on the charge trajectory. We have here a kind of {\it dynamical classical nonlocality} attributable to the equations of electrodynamics. We can alternatively see that the momentum $\mathbf{L}_\texttt{AB}= q\Phi\hat{\mathbf{z}}/(2\pi c)$ is a nonlocal quantity by writing it in terms of the flux $\Phi=\pi R^2 B$. The equivalent relation reads
\begin{align}
\mathbf{L}_{\texttt{AB}}= \frac{nq R^2}{2c} \mathbf{B}.
\end{align}
Since the field $\mathbf{B}$ is zero everywhere outside the solenoid, the region where the charge $q$ is moving, it follows that there is no Lorentz force locally acting on the charge and, however, there is a nonlocal action of the field $\mathbf{B}$ on the charge $q$, which is manifested by the existence of the momentum $\mathbf{L}_{\texttt{AB}}$. It is interesting to note that the integration limits in Eq.~(16) explicitly exclude the region where the magnetic field is non-zero, and however, after the integration process the magnetic field reappears as seen in Eq.~(22). This is the ``magic" of nonlocality.

\section{Alice, Bob and the nonlocality of the  momentum {\large $\mathbf{\texttt{L}}_\texttt{AB}$}}\label{V}
\noindent Alice and Bob discuss about the physical origin of the electromagnetic angular momentum in a region where there is no magnetic field, but there is a Coulomb-gauge vector potential and a localised electric charge density. 
While Bob thinks that this momentum is due to the action of the vector potential on the charge density occurring in the volume, Alice is convinced that this momentum is originated by the interaction of the potential and the electric field due to the charge density occurring in the surface of that volume.

Bob argues: ``In order to find the momentum, I can apply my formula
\begin{align}
\textbf{L}=\frac{1}{c}\int_{V}\mathbf{x} \times (\rho_\texttt{e} \mathbf{A})\,d^3x.
\end{align}
As you can see Alice, I require the values of the electric charge density $\rho_\texttt{e}$ and the vector potential $\mathbf{A}$ inside the volume \emph{V} to determine the momentum $\textbf{L}.$ After performing the volume integration, the momentum is generally a function of points in the volume.
Therefore, I conclude that this momentum is a local effect of the interaction of the vector potential on the charge density produced in the volume.'' In short: for Bob the momentum is a local effect happening in the volume.

Alice then argues: ``In order to find the momentum, I can use my formula
\begin{align}
\textbf{L}\!=\!-\frac{1}{4\pi c}\!\oint_S \mathbf{x} \!\times\!\big[\hat{\textbf{n}}(\textbf{E}\!\cdot\! \textbf{A})\!-\!\textbf{A}(\hat{\textbf{n}}\cdot\textbf{E})\!-\!\textbf{E}(\hat{\textbf{n}}\!\cdot\! \textbf{A})\big]d{S}.
\end{align}
As you can see Bob, I only require the values of the electric field \textbf{E} and the potential \textbf{A} on the surface $S$
to determine the momentum $\textbf{L}$. After performing the surface integration, the momentum is generally a function of points in the surface. But the localised charge density is not defined in that surface. Therefore, I conclude that this momentum is originated by the interaction of the electric field and the vector potential occurring in the surface. This momentum is nonlocal with respect to the charge density because the latter is defined in the volume.''  In a few words: for Alice the momentum is a nonlocal effect originated in the surface.

With the idea of applying their respective formulas in a specific case, Alice and Bob choose the electromagnetic AB configuration in which an electric charge $q$ is moving around an infinitely-long solenoid of radius $R$ enclosing the magnetic flux $\Phi=\pi R^2 B$, where $B$ is a uniform magnetic field. They both know that this configuration is defined in a non-simply connected region.

Using the AB potential and the charge density, Bob's formula gives the momentum $\textbf{L}=qR^2 \mathbf{B}/(2c)$.
He is surprised because this momentum cannot be interpreted as a local effect of the vector potential produced in the volume. He claims: ``The existence of the momentum seems to be a consequence of the nonlocal action of the magnetic field of the solenoid on the electric charge because the latter is moving in a region where the former is excluded. This an unusual prediction of the equations of classical electrodynamics. I knew that the magnetic field may produce nonlocal effects in quantum mechanics but not in classical mechanics!''

Using the AB potential and the electric field of the point charge, Alice's formula also gives the momentum $\textbf{L}=qR^2\mathbf{B}/(2c)$. She is astonished because this momentum cannot be interpreted as a nonlocal effect of the interaction of the vector potential and the electric field occurring in the surface. The momentum seems to be caused by the confined magnetic field of the solenoid acting on the electric charge at a distance. She says: ``It seems to be that the magnetic field communicates `telepathically' with the moving charge, without any carrier of the interaction travelling between them! The nonlocality is originated in the solenoid and not in the surface as I believed. Accordingly, the momentum is a nonlocal manifestation of the magnetic field.''

From the decomposition given in Eq.~(11), it becomes clear that Bob's formula is equivalent to  Alice's formula. To see how Alice's formula leads to the momentum $\textbf{L}=q\Phi\hat{\mathbf{z}}/(2\pi c)=qR^2 \textbf{B}/(2c)$, see the Appendix C.

The existence of the nonlocal momentum $\textbf{L}_\texttt{AB}$ breaks the common misconception that the magnetic field may produce nonlocal effects only in quantum mechanics but not in classical mechanics and that this represents a clear difference between classical and quantum theories.\cite{21} As far as the electromagnetic AB configuration is concerned, the difference between both theories is with respect to the predicted effects. While quantum mechanics predicts the existence of the accumulated phase $\delta_\texttt{AB}=nq\Phi/(\hbar c)$, classical mechanics predicts the existence of the accumulated momentum $L_\texttt{AB}= nq\Phi/(2\pi c)$.

\section{Gauge Invariance of the momentum {\large $\mathbf{\texttt{L}}_\texttt{AB}$}}\label{VI}
\noindent
If we apply a gauge transformation $ \mathbf{A}'= \mathbf{A}+ \gradv\Lambda$ to the right-hand side of Eq.~(13) we obtain
\begin{align}
\nonumber \frac{1}{c}\int_{V}\mathbf{x} \times(\rho_\texttt{e} \mathbf{A'})\,d^3x=&\, \frac{1}{c} \int_{V}\mathbf{x}  \times  (\rho_\texttt{e} \mathbf{A})\,d^3x \\
&\,\,+\frac{1}{c}\int_{V}\mathbf{x}\times (\rho_\texttt{e} \gradv\Lambda)\,d^3x.
\end{align}
The Maxwell formula in Eq.~(13) will be gauge invariant if the last term of Eq.~(25) vanishes. To show that this is indeed the case in the electromagnetic AB configuration we must to take into account four aspects: (i) Since the AB potential is in the Coulomb gauge $\gradv \cdot \textbf{A}= 0$ it follows that a further gauge transformation applied to this potential must be a \emph{restricted} gauge transformation, that is, one in which the gauge function $\Lambda$ satisfies the Laplace equation $\gradv^2\Lambda=0.$ (ii) In the electromagnetic AB configuration the charged particle moves in the $x$-$y$ plane. Therefore the restricted gauge function in this two-dimensional space will be of the generic form $\Lambda = \Lambda (\rho,\phi).$ (iii) The gauge function $\Lambda$ must be a single-valued function: $\Lambda(\rho,\phi)=\Lambda(\rho, \phi +2\pi)$ and its gradient $\gradv\Lambda$ must be a non-singular function. (iv) The general solution of the Laplace equation in polar coordinates
\begin{align}
\gradv^2\Lambda=\frac{\partial^2 \Lambda}{\partial \rho^2} +\frac{1}{\rho} \frac{\partial \Lambda}{\partial \rho} + \frac{1}{\rho^2} \frac{\partial^2 \Lambda}{\partial \phi^2}=0,
\end{align}
for a gauge function $\Lambda$ satisfying the properties specified in (iii) is given by\cite{15}
\begin{align}
\Lambda(\rho,\phi)= \sum^{\infty}_{m=1}\,\rho^{m}\big(\alpha_m\cos(m\phi)+\beta_m \sin(m\phi)\big),
\end{align}
where $m$ is a positive integer and $\alpha_m$ and $\beta_m$ are the constant expansion coefficients given by
\begin{align}
\alpha_{m} =& \,\frac{1}{\rho^{m}\pi} \int_{0}^{2\pi}\! \Lambda(\rho, \phi)\,\cos(m\phi)\,d\phi\,,\\
\beta_{m}=& \,\frac{1}{\rho^{m}\pi} \int_{0}^{2\pi} \!\Lambda(\rho, \phi)\,\sin(m\phi)\,d\phi\,.
\end{align}
Having in mind the above aspects, let us write the last term of Eq.~(25) as $(1/c)\int_{V}\mathbf{x}'\times[\rho_\texttt{e}(\textbf{x}') \gradv'\Lambda (\textbf{x}')]\,d^3x'$. Inserting $\textbf{x}' = \rho' \hat{\bfrho}, \rho_\texttt{e}(\textbf{x}')= q\delta(\rho' -\rho)\delta(z')/(2\pi \rho')$ and  $\gradv'\Lambda(\textbf{x}')=(\partial \Lambda/\partial \rho') \hat{\bfrho}+ (1/\rho')(\partial \Lambda/\partial \phi')\hat{\bfphi}$, we obtain
\begin{align}
 \nonumber\frac{1}{c}\int_{V}\mathbf{x}'&\times\big[\rho_\texttt{e}(\textbf{x}') \gradv'\Lambda (\textbf{x}')\big]\,d^3x'
 \\ \nonumber=& \,\frac{q \,\hat{\textbf{z}}}{2 \pi c}\int^{+\infty}_{-\infty}\!\!\delta(z')\,dz'\int_{R}^{\infty} \!\delta(\rho' -\rho)\,d\rho' \oint \frac{\partial \Lambda}{\partial \phi'}\,d\phi'\\
 =&\, \frac{q \Theta(\rho-R)\hat{\textbf{z}}}{2 \pi c}\oint\! \frac{\partial \Lambda}{\partial \phi'}d\phi'=\frac{q \hat{\textbf{z}}}{2 \pi c}\!\oint \!\frac{\partial \Lambda}{\partial \phi'}\,d\phi',
\end{align}
where $\Theta(\rho-R)=1$ because the charge moves outside the solenoid ($\rho>R).$ To show that the integral $\oint [\partial \Lambda/\partial \phi']d\phi'$ vanishes we insert Eq.~(27) in this integral, obtaining
\begin{align}
\nonumber \oint& \frac{\partial \Lambda}{\partial \phi'}\, d\phi' \\
\nonumber=&  \oint\! \sum^{\infty}_{m=1} \frac{\partial }{\partial \phi'}\bigg[\rho'^{\,m}\big(\alpha_m\cos(m\phi')\!+\!\beta_m \sin(m\phi')\big)\bigg]d\phi' \\
=& \!\sum^{\infty}_{m=1} \!m\rho'^{\,m}\!\!\oint\! \big( \beta_m \cos(m\phi')\!-\! \alpha_m \sin(m\phi')\big)d\phi' \!=\!0,
\end{align}
where we have used the fact that $\oint \cos(m \phi')d\phi' = 0$ and $\oint \sin(m \phi')d\phi'=0$ when $m$ is a positive integer.
Equations (30) and (31) imply the vanishing of the term $(1/c)\int_{V}\mathbf{x}'\times[\rho_\texttt{e}(\textbf{x}') \gradv'\Lambda (\textbf{x}')]\,d^3x'$.
This shows that the Maxwell formula in Eq.~(13) is gauge invariant in the case of the electromagnetic AB configuration.

\section{Physical manifestation of the momentum {\large $\mathbf{\texttt{L}}_\texttt{AB}$}}\label{VII}
\noindent The momentum $L_{\texttt{AB}} = q \Phi/(2 \pi c)$ implies $\Phi=2\pi c L_{\texttt{AB}}/q$ which combines with Eq.~(9) to give
\begin{align}
\mathbf{A} = \frac{cL_\texttt{AB}}{q\rho} \hat{\bfphi}.
\end{align}
This relation will be useful to answer the next question: How can we make the momentum $\mathbf{L}_{\texttt{AB}}$ manifest physically? Let us ignore retardation effects and therefore the subsequent considerations belong to the quasi-static regime. If we turn off the current in the solenoid then the magnetic flux varies with time $\Phi(t)=\pi R^2B(t)$ during a short period of time. This process implies that the momentum $L_{\texttt{AB}}$ instantaneously varies with time:  $L_{\texttt{AB}}(t) = q \Phi(t)/(2 \pi c)$. Therefore the potential in Eq.~(32) varies with time: $\mathbf{A}(t) = cL_\texttt{AB}(t)\hat{\bfphi}/(q\rho)$ and then an electric field $\textbf{E}=-(1/c)d\textbf{A}/dt$ is induced, which exerts a force on the particle of mass $m$ and charge $q$ which is assumed to move in the $x$-$y$ plane.  The field $\mathbf{E}$ reads
\begin{align}
\mathbf{E} =- \frac{d}{dt}\bigg(\frac{L_\texttt{AB}}{q\rho} \hat{\bfphi}\bigg).
\end{align}
We note that the mechanical force exerted on the charged particle is $d \mathbf{p}/dt= q \mathbf{E}$ where $\mathbf{p}=m\textbf{\textit{v}}$ is its mechanical momentum. If $\mathbf{x}$ is a fixed position vector then the following basic relation follows: $\mathbf{x} \times d \mathbf{p}/dt=\mathbf{x} \times q \mathbf{E}$. We now examine both sides of this relation. The left-had side can be written as $\mathbf{x} \times d \mathbf{p}/dt=d (\mathbf{x} \times\mathbf{p})/dt=d \mathbf{L}_{\texttt{mech}}/dt$, where $\mathbf{L}_{\texttt{mech}}=\mathbf{x}\times\mathbf{p}$ is the mechanical angular momentum.  If now we write $\mathbf{x} = \rho \hat{\bfrho}$ and use Eq.~(33) then the right-hand side of the basic relation becomes $\mathbf{x} \times q \mathbf{E}= -d \mathbf{L}_{\texttt{AB}}/dt$. Equating both sides of the basic equation we get the conservation law of the total angular momentum
\begin{align}
\frac{d}{dt}\big( \mathbf{L}_{\texttt{mech}}+\mathbf{L}_{\texttt{AB}}\big)=0.
\end{align}
Hence, $\mathbf{L}_{\texttt{AB}}=-\mathbf{L}_{\texttt{mech}}+$ constant, which shows that the existence of the momentum $\mathbf{L}_{\texttt{AB}}$ (previously stored in the electromagnetic  AB configuration) may be revealed through the existence of the mechanical angular momentum $\mathbf{L}_{\texttt{mech}}$.

It is interesting to note that we could eventually measure the momentum $L_{\texttt{AB}}$ with the double-slit experiment used to detect the AB phase. According to this experiment, an enclosed magnetic flux due to an infinitely-long solenoid is placed between two interfering beams of charged particles in a double-slit arrangement and as a result the whole double-slit interference pattern is shifted by the amount\cite{20}
\begin{align}
\Delta x = \frac{l\barlambda}{d}\delta_{\texttt{AB}}\,,
\end{align}
where $d$ is the separation between the two slits on the first screen, $l$ is the distance between the two screens, $\barlambda= \hbar/mv$ is the reduced de Broglie wavelength of the charges $q,$ with mass $m$ and velocity $v,$ and $\delta_{\texttt{AB}} = q\Phi/\hbar c$ is the AB phase with winding number $n=1.$
Inserting  $\Phi=\hbar c L_{\texttt{AB}}/q$ and $\barlambda=  \hbar/ mv$ in Eq.~(35), we obtain
\begin{align}
\Delta x = \frac{2 \pi l}{mv d}\,L_\texttt{AB}\,,
\end{align}
which shows that the shifted interference pattern is determined by the electromagnetic angular momentum $L_\texttt{AB}$ and the linear mechanical momentum $mv$. It is interesting to note that there is no explicit quantum signature in the shift given by Eq.~(36) [$\hbar$ has disappeared]. This opens the door to the speculative idea that the shift $\Delta x$ could be explained using classical considerations in which the momentum $L_\texttt{AB}$ might play a central role.

Using Eq.~(36) we can obtain an expression for the momentum $L_\texttt{AB}$ in terms of measurable quantities:
\begin{align}
L_\texttt{AB}= \frac{\Delta x\, mv d}{2 \pi \,l}.
\end{align}
Therefore the classical momentum $L_\texttt{AB}$ could be measured using the quantum-mechanical AB set-up.

\section{Discussion}\label{VIII}
\noindent We can combine the momentum $L_\texttt{AB}$ and the phase $\delta_\texttt{AB}$ to obtain the linear relation  $\delta_\texttt{AB}=(2\pi/\hbar)L_\texttt{AB},$ which is of conceptual importance since it emphasizes a natural connection between predictions of quantum and classical physics. Let us obtain this relation using two different starting expressions. From the AB phase expressed as
 $\delta_\texttt{AB}=[q/(\hbar c)]\oint \mathbf{A}\cdot d \mathbf{l}$, the AB potential given in Eq.~(32) and $d \textbf{l}=\rho d\phi\hat{\bfphi}$  we get the desired relation
\begin{align}
\delta_\texttt{AB}= \frac{q}{\hbar c}\oint \mathbf{A}\cdot d \mathbf{l}=\frac{2\pi}{\hbar}L_\texttt{AB}.
\end{align}
On the other hand, using $\Phi=\hbar c\delta_\texttt{AB}/q$ in Eq.~(32) we obtain the alternative form
\begin{align}
\mathbf{A} = \frac{\hbar c\delta_\texttt{AB}\hat{\bfphi}}{2\pi q\rho}.
\end{align}
We can write the magnetic flux as $\Phi=\oint \mathbf{A}\cdot d \mathbf{l}$ and use it in $L_{\texttt{AB}}=q\Phi/(2\pi c)$ to obtain  $L_{\texttt{AB}}=[q/(2\pi c)]\oint \mathbf{A}\cdot d \mathbf{l}$. This expression and Eq.~(39) together with $d \textbf{l}=\rho d\phi\hat{\bfphi}$  yield again the expected relation
\begin{align}
L_{\texttt{AB}} =  \frac{q}{2\pi c}\oint\mathbf{A}\cdot d \mathbf{l}=\frac{\hbar}{2\pi}\delta_\texttt{AB}.
\end{align}
We then conclude that the momentum $L_{\texttt{AB}}$ and the phase $\delta_\texttt{AB}$ are related to the extent that the former is the classical counterpart of the latter, or equivalently, the latter is the quantum mechanical counterpart of the former. This conclusion is consistent with the fact that
both the momentum $L_{\texttt{AB}}$ and the phase $\delta_\texttt{AB}$ share the same electromagnetic configuration.

\section{Conclusion}\label{IX}
\noindent We return now to the questions (i)-(vii) raised in our introduction.

\begin{enumerate}[label=(\roman*)]
  \item Can the Poynting formula given in Eq.~(4) be used to obtain the momentum $\mathbf{L}_\texttt{AB}$? \emph{Answer}: Yes, whenever this formula be decomposed into its volume and surface parts. The volume part gives the momentum
$\mathbf{L}_\texttt{AB}$ and the surface part yields the return momentum $-\mathbf{L}_\texttt{AB}$. Both contributions imply that the momentum given by the Poynting formula is zero outside the solenoid. But there is no physical inconsistence because the return momentum  is defined in a different region (the infinite surface) to that where the charge $q$  is moving (the volumetric space outside the solenoid).

  \item How are explicitly related the Maxwell and Poynting formulas given in Eqs.~(3) and (4)? \emph{Answer}: They are related by means of Eq.~(11). This novel equation is derived from the identity in Eq.~(10) which is in turn demonstrated in Appendix B.

\item  Is the Maxwell formula in Eq.~(3) gauge invariant? \emph{Answer}: Yes, the gauge-invariance of Eq.~(3) is demonstrated in Sec.~\ref{VI}.

\item Does the momentum $\mathbf{L}_\texttt{AB}$ share the topological and nonlocal features of the AB phase $\delta_\texttt{AB}$? \emph{Answer}: Yes, like the phase $\delta_\texttt{AB}$, the momentum $\mathbf{L}_\texttt{AB}$ is topological because it depends on the number of windings the electric charge carries out around the solenoid and is independent of the shape of the charge motion. This allows us to introduce the idea of an accumulated electromagnetic angular momentum. Like the phase $\delta_\texttt{AB}$, the momentum $\mathbf{L}_\texttt{AB}$ is nonlocal because the magnetic flux has no local consequences at any point on the charge trajectory.

\item Can the momentum $\mathbf{L}_\texttt{AB}$ be considered the classical counterpart of the AB phase $\delta_\texttt{AB}$? \emph{Answer}: Yes, and this answer is supported by their linear relation $\delta_\texttt{AB}=(2\pi/\hbar) L_\texttt{AB}$. With equal right we could say that the AB phase $\delta_\texttt{AB}$ is the quantum mechanical counterpart of the momentum $\mathbf{L}_\texttt{AB}$.

\item Is it crucial for the momentum $\mathbf{L}_\texttt{AB}$ that the AB configuration is defined in a non-simply connected region?
\emph{Answer}: Yes, if for example, the charge would not surround the solenoid then this configuration would be in a simply-connected region outside the solenoid. In this case the electromagnetic angular momentum would vanish.

\item Could we conclude that the momentum $\mathbf{L}_\texttt{AB}$ is a nonlocal manifestation of the magnetic field in classical physics? \emph{Answer}: Yes, and this conclusion opens the door to the idea of a dynamical nonlocality attributable to equations of electrodynamics.
\end{enumerate}

Nonlocality is often portrayed as being equivalent to a spooky action at a distance.\cite{22} For this reason many physicists are reluctant to accept that nonlocality is a feature of nature. But now besides the dynamical quantum nonlocality\cite{3,4} manifested in the AB phase $\delta_\texttt{AB}$ or the kinematic quantum nonlocality manifested in the Bell-inequality violations,\cite{3} we also have the dynamical classical nonlocality manifested in the  momentum $L_\texttt{AB}$. This encourages us to look for more nonlocal manifestations in physics. Wilczek has launched the challenging question:\cite{23} ``What will the next 100 years in physics bring?" We have good reasons to think that within 100 years we will have many nonlocal physical theories.

\appendix
\section*{A. Derivation of Eq.~(8) and its regularisation}
\renewcommand\theequation{A\arabic{equation}}
\setcounter{equation}{0}

\noindent Our starting point is the Poisson equation for the vector potential given in Eq.~(7). This equation can be written as $\gradv^2\textbf{A}=-4\pi \gradv\times \textbf{M}$, where $\textbf{M}$ is the magnetisation vector $\textbf{M}=\Phi\Theta(R-\rho)\hat{\textbf{z}}/(4\pi^2 R^2)$. The solution reads
\begin{align}
\textbf{A}(\textbf{x})=\int \frac{ \gradv' \times \textbf{M} (\textbf{x}')}{|\textbf{x}-\textbf{x}'|}\,d^3x',
\end{align}
where $\textbf{x}$ is the field point and $\textbf{x}'$ the source point and the integration is over all space. Using $(\gradv'\times \textbf{M})/|\textbf{x} -\textbf{x}'|= \gradv \times (\textbf{M}/|\textbf{x} -\textbf{x}'|)+\gradv' \times (\textbf{M}/|\textbf{x} -\textbf{x}'|),$ Eq.~(A1) becomes
\begin{align}
\textbf{A}(\textbf{x})= \!\!\int\!\gradv \times \bigg(\frac{\textbf{M} (\textbf{x}')}{|\textbf{x}\!-\!\textbf{x}'|}\bigg)d^3x'\!+\! \int\!\gradv' \times \bigg(\frac{ \textbf{M} (\textbf{x}')}{|\textbf{x}\!-\!\textbf{x}'|}\bigg)d^3x'.
\end{align}
Using $\gradv \times (\textbf{M}(\textbf{x}')/|\textbf{x} -\textbf{x}'|)= \textbf{M}(\textbf{x}')\times (\textbf{x} -\textbf{x}')/|\textbf{x} -\textbf{x}'|^3,$  and transforming the second volume integral of Eq.~(A2) into a surface integral we obtain
\begin{align}
\textbf{A}(\textbf{x}) = \!\int\! \frac{\textbf{M}(\textbf{x}') \times (\textbf{x} \!-\!\textbf{x}')}{|\textbf{x}\! -\!\textbf{x}'|^3}d^3x'\!+\!\oint \!\hat{\textbf{n}} \times \bigg( \frac{\textbf{M}(\textbf{x}')}{|\textbf{x}\!-\!\textbf{x}'|} \bigg)dS',
\end{align}
where $\hat{\textbf{n}}$ is the outward vector to the surface enclosing the infinite volume.  The magnetisation $\textbf{M}$ is confined in a finite region of space and this means that it is zero outside this region, while the surface integral in the above expression encloses all space. Therefore the magnetisation vanishes at every point on the infinite surface and the surface integral in Eq.~(A3) vanishes. In our case the magnetisation is given by $\textbf{M} = \Phi \Theta(R\!-\!\rho)\hat{\textbf{z}}/(4 \pi^2 R^2)$. It is clear that $\textbf{M}=0$  as $\rho \rightarrow \infty$ because $\Theta(R\!-\!\rho)=0$ as $\rho \rightarrow \infty.$ Hence Eq.~(A3) reduces to
\begin{align}
\textbf{A}(\textbf{x}) = \int \frac{\textbf{M}(\textbf{x}') \times (\textbf{x} -\textbf{x}')}{|\textbf{x} -\textbf{x}'|^3}\,d^3x'.
\end{align}
Now, the separation vector is given by
\begin{align}
\nonumber \textbf{x} \!-\!\textbf{x}'=&\frac{\big(\rho^2 \!-\! \rho'(x \!\cos\!\phi' \!+\! y\!\sin\!\phi')\big)\hat{\bfrho}\!+\!\rho'\big(y\!\cos\!\phi'\!-\!x\!\sin\!\phi' \big)\hat{\bfphi}}{\rho}
\\ &+(z-z')\hat{\textbf{z}}\,,
\end{align}
and its corresponding magnitude gives
\begin{align}
|\textbf{x} \!-\!\textbf{x}'|=& \sqrt{\rho^2\!+\!\rho'^{2}\!-\!2\rho'(x\cos\phi' \!+\! y\sin\phi')+(z\!-\!z')^2}\,.
\end{align}
Inserting Eqs.~(A5) and (A6) and the magnetisation vector $\textbf{M} = \Phi \Theta(R\!-\!\rho)\hat{\textbf{z}}/(4 \pi^2 R^2)$ in Eq.~(A4) we obtain
\begin{align}
\nonumber&\textbf{A}=\frac{\Phi}{(2\pi R)^2}\\
\nonumber\times&\Bigg[\!\int \!\frac{\hat{\bfphi}\,\Theta(R \!-\! \rho')\big( \rho^2 \!-\! \rho'(x\cos\phi'\!+\!y\sin\phi') \big) \rho'd\rho'd\phi'dz'}{\rho\big(\rho^2\!+\!\rho'^{2}\!-\!2\rho'(x\cos\phi'\!+\! y\sin\phi')\!+\!(z\!-\!z')^2 \big)^{3/2}} \\
&+ \!\int\!\!\frac{\hat{\bfrho}\,\big[\rho'(x\sin\phi' \!-\!y\cos\phi')\big]\,\rho'd\rho'd\phi'dz'}{\rho\big( \rho^2\!+\!\rho'^{2}\!-\!2\rho'(x\cos\phi' \!+\! y\sin\phi')\!+\!(z\!-\!z')^2 \big)^{3/2}} \Bigg].
\end{align}
The $z'$ integrals are straightforward to evaluate via a change of variable. It follows
\begin{align}
\nonumber\int_{-\infty}^{+\infty}  &\frac{dz'}{\big( \rho^2\!+\!\rho'^{2}\!- 2\rho(x\cos\phi' \!+\! y\sin\phi')\!+\!(z\!-\!z')^2 \big)^{3/2}}\\
=& \,\frac{2}{\big( \rho^2+\rho'^{2}-2\rho'(x\cos\phi' + y\sin\phi')\big)}\,.
\end{align}
Using Eq.~(A8) in Eq.~(A7) we obtain
\begin{align}
\nonumber\textbf{A}&=\, \frac{\Phi}{2\pi^2R^2}\int^{\rho}_{0} \rho'\,\Theta(R\!-\!\rho')\,d\rho' \\
\nonumber\times&\Bigg[\int_{0}^{2\pi}\!\! \frac{\hat{\bfphi}\,\big( \rho^2 \!-\! \rho'(x\cos\phi'\!+\!y\sin\phi')\big)\,d\phi'}{\rho\,\big( \rho^2+\rho'^{2}-2\rho'\,(x\cos\phi' + y\sin\phi')\big)}\\
&+ \int_{0}^{2\pi} \!\!\frac{\hat{\bfrho}\,\rho'(x\sin\phi' \!-\!y\cos\phi')\,d\phi'}{\rho\,\big( \rho^2+\rho'^{2}-2\rho'\,(x\cos\phi' + y\sin\phi')\big)}\Bigg].
\end{align}
The $\phi'$ integrals can be evaluated via contour integrations in the complex plane. We obtain the results
\begin{align}
\int_{0}^{2\pi} \!\!\frac{\big(\rho^2 - \rho'(x\cos\phi'+y\sin\phi')\big)\,d\phi'}{\rho\,\big( \rho^2+\rho'^{2}-2\rho'(x\cos\phi' + y\sin\phi')\big)}=& \frac{2\pi}{\rho}\,,\\
\int_{0}^{2\pi}\!\frac{\rho'(x\sin\phi' -y\cos\phi')\,d\phi'}{\rho\,\big( \rho^2+\rho'^{2}-2\rho'(x\cos\phi' + y\sin\phi')\big)}\, =\,& 0\,,
\end{align}
where we have assumed the condition $\rho^2>\rho'^2>0.$ Inserting Eqs.~(A10) and (A11) in Eq.~(A9) we obtain
\begin{align}
\textbf{A}= \frac{\Phi\,\hat{\bfphi}}{\pi R^2\rho}\int^{\rho}_{0} \rho'\,\Theta(R-\rho')\,d\rho'.
\end{align}
Now, the $\rho'$ integral gives
\begin{align}
\nonumber \int^{\rho}_{0}\rho'\, &\Theta(R-\rho')\,d\rho'\\
\nonumber=&\, \frac{1}{2}\bigg(R^2\,\Theta(\rho-R)+\rho^2\,\big(1-\Theta(\rho-R) \big) \bigg)\\
=&\, \frac{1}{2}\bigg( R^2\,\Theta(\rho-R)+\rho^2\,\Theta(R-\rho) \bigg),
\end{align}
where in the last step we used the property $\Theta(-x)=1-\Theta(x).$ Inserting Eq.~(A13) in Eq.~(A12) we obtain
\begin{align}
\textbf{A}=\bigg(\frac{\Phi}{2\pi\rho}\Theta(\rho-R)+\frac{\Phi\rho}{2\pi R^2}\Theta(R-\rho)\bigg)\hat{\bfphi},
\end{align}
which is the potential in Eq.~(8).

In order to show the potential in Eq.~(8) is continuous at $\rho=R$ we need to regularise it.  An appropriate regularisation of Eq.~(8) can be obtained by making the replacement $\rho\rightarrow \rho + \epsilon,$ where $\epsilon>0$ is an infinitesimal quantity. The regularised potential is
\begin{align}
\textbf{A}_{\epsilon} = \bigg( \frac{\Phi\Theta[(\rho\!+\!\epsilon)\!-\!R]}{2\pi(\rho\!+\!\epsilon)} + \frac{\Phi(\rho\!+\!\epsilon)\,\Theta[R\!-\!(\rho\!+\!\epsilon)]}{2\pi R^2}\bigg)\hat{\bfphi}.
\end{align}
Clearly, in the limit $\epsilon \rightarrow 0$ we recover Eq.~(8). At $\rho = R$ the regularised potential in Eq.~(A15) reads
\begin{align}
\textbf{A}_{\epsilon}(R) =\bigg( \frac{\Phi\Theta[+\epsilon]}{2\pi(R+\epsilon)} + \frac{\Phi(R+\epsilon)\,\Theta[-\epsilon]}{2\pi R^2}\bigg)\hat{\bfphi}.
\end{align}
In the limit $\epsilon \rightarrow 0$ we obtain
\begin{align}
\nonumber \lim_{\epsilon\rightarrow 0}\,\textbf{A}_{\epsilon}(R) =&\, \bigg( \frac{\Phi\Theta[+\epsilon]}{2\pi(R+\epsilon)} + \frac{\Phi(R+\epsilon)\,\Theta[-\epsilon]}{2\pi R^2}\bigg)\hat{\bfphi}\\
=& \,\frac{\Phi}{2\pi}\frac{\hat{\bfphi}}{R}\,,
\end{align}
where we have used the results $\Theta[+\epsilon]=1$ and $\Theta[-\epsilon]=0$ as $\epsilon\rightarrow 0.$ Therefore the potential in Eq.~(8) is continuous at $\rho=R$ and has the form $\textbf{A}(R) = \Phi \hat{\bfphi}/(2 \pi R).$

\section*{B. Verification of Eq.~(10)}
\renewcommand\theequation{B\arabic{equation}}
\setcounter{equation}{0}
\noindent A direct operation on the left-hand side of Eq.~(10) gives
\begin{align}
\nonumber [\mathbf{x} \times(\mathbf{E}\times \mathbf{B})]^\texttt{s}=&\,\varepsilon^{\texttt{sqi}}x_\texttt{q}\big(\mathbf{E}\times \mathbf{B}\big)_\texttt{i}\nonumber\\=&\,
\nonumber\varepsilon^{\texttt{sqi}}x_\texttt{q}\big(\varepsilon_{\texttt{ikp}}E^{\texttt{k}}B^\texttt{p}\big)\\
=&\, \varepsilon^{\texttt{sqi}}x_\texttt{q}E^\texttt{k}\big(\varepsilon_{\texttt{pik}}B^\texttt{p}\big).
\end{align}
Now, using $B^\texttt{p}=\varepsilon^{\texttt{plm}}\partial_\texttt{l}A_\texttt{m}$ it follows
\begin{align}
\nonumber\varepsilon_{\texttt{pik}}B^\texttt{p}=&\, \varepsilon_{\texttt{pik}}\varepsilon^{\texttt{plm}}(\partial_\texttt{l}A_\texttt{m})\\
\nonumber = &\,\big(\delta^{\texttt{l}}_{i}\delta^{\texttt{m}}_{\texttt{k}}-\delta^{\texttt{m}}_{\texttt{i}}\delta^{\texttt{l}}_{\texttt{k}}\big)\partial_\texttt{l}A_\texttt{m}\\
=&\,\partial_\texttt{i}A_\texttt{k}-\partial_\texttt{k}A_\texttt{i},
\end{align}
where we have used the identity $\varepsilon_{\texttt{pik}}\varepsilon^{\texttt{plm}}=\delta^{\texttt{l}}_{i}\delta^{\texttt{m}}_{\texttt{k}}-\delta^{\texttt{m}}_{\texttt{i}}\delta^{\texttt{l}}_{\texttt{k}}.$ Using Eq.~(B2) in Eq.~(B1) we obtain
\begin{align}
[\mathbf{x} \times(\mathbf{E}\times \mathbf{B})]^\texttt{s}
=\, \varepsilon^{\texttt{sqi}}x_\texttt{q}E^\texttt{k}\big( \partial_\texttt{i}A_\texttt{k}-\partial_\texttt{k}A_\texttt{i} \big).
\end{align}
Now, let us show the following result
\begin{align}
\nonumber E^\texttt{k}\big(\partial_\texttt{i}A_\texttt{k}\!-\!\partial_\texttt{k}A_\texttt{i}\big)=&\, 4\pi\rho_\texttt{e}A_\texttt{i}\\&+
\partial^\texttt{k}\big(\delta_\texttt{ik}E_\texttt{m}A^\texttt{m}\!-\!E_\texttt{k}A_\texttt{i}\!-\!E_\texttt{i}A_\texttt{k}\big).
\end{align}
A direct calculation on the second term of Eq.~(B4) gives
\begin{align}
\nonumber \partial^\texttt{k}\big(\delta_\texttt{ik}E_\texttt{m}A^\texttt{m}\!&-\!E_\texttt{k}A_\texttt{i}\!-\!E_\texttt{i}A_\texttt{k}\big)\\=
\nonumber &\,A^\texttt{k}\big(\partial_\texttt{i}E_\texttt{k}\!-\!\partial_\texttt{k}E_\texttt{i}\big)
\!+\!E^\texttt{k}\big(\partial_\texttt{i}A_\texttt{k}\!-\!\partial_\texttt{k}A_\texttt{i}\big)
\\&\,-E_\texttt{i}\partial^\texttt{k}A_\texttt{k}-A_\texttt{i}\partial^\texttt{k}E_\texttt{k}.
\end{align}
Using $\partial_\texttt{i}E_\texttt{k}-\partial_\texttt{k}E_\texttt{i}=\varepsilon_{\texttt{pik}}(\varepsilon^{\texttt{plm}}\partial_\texttt{l}E_\texttt{m})=0$ ($\gradv\times \textbf{E}=0$ for electrostatic fields), $\partial^\texttt{k}A_\texttt{k}=0$ (Coulomb gauge) and $\partial^\texttt{k}E_\texttt{k}=4\pi\rho_\texttt{e}$ (Gauss' law) in Eq.~(B5) we obtain an equation that implies Eq.~(B4). Inserting Eq.~(B4) into Eq.~(B3) we finally obtain Eq.~(10).

\section*{C. Derivation of the return momentum {\large $\mathbf{\texttt{L}}^{\texttt{S}}_\texttt{AB}$}}~\label{C}
\renewcommand\theequation{C\arabic{equation}}

\setcounter{equation}{0}
\noindent
According to Eq.~(14), the return momentum for the AB configuration can be calculated using the expression
\begin{align}
\nonumber\textbf{L}^\texttt{S}_{\texttt{AB}}\!=\!\frac{1}{4\pi c}\oint_S \mathbf{x}'\times \big[\hat{\textbf{n}}&\big(\textbf{E}(\mathbf{x}')\!\cdot\! \textbf{A}(\mathbf{x}')\big)\!-\!\textbf{A}(\mathbf{x}')\big(\hat{\textbf{n}}\! \cdot\! \textbf{E}(\mathbf{x}')\big)\!\\
&-\!\textbf{E}(\mathbf{x}')\big(\hat{\textbf{n}}\!\cdot\! \textbf{A}(\mathbf{x}')\big)\big] d{S}'.
\end{align}
Here we assume the charge $q$ is located at the generic point $\textbf{x}= x\hat{\textbf{x}} + y\hat{\textbf{y}}$
where $\sqrt{x^2+y^2}>R,$ with $R$ being the solenoid radius and $\sqrt{x^2+y^2}=\rho.$ We also note that the surface integral in Eq.~(C1) must be evaluated in the limit $\rho' \rightarrow \infty,$ which must be taken at the end of the integration process. This is consistent with the fact that we have chosen a cylindrical surface of infinite radius as the surface for which $\textbf{L}^\texttt{S}_{\texttt{AB}}$ must be evaluated.

The electric field $\textbf{E}(\textbf{x}') = q(\textbf{x}'-\textbf{x})/|\textbf{x}'-\textbf{x}|^3$ produced by the charge $q$ located at $\textbf{x}$ is given by
\begin{align}
\textbf{E}\!=\!\frac{q\big[\big( \rho' \!-\! (x\!\cos\!\phi'\!+\!y\!\sin\!\phi')\!\big)\hat{\bfrho}\!+\!(x\!\sin\!\phi'\!-\!y\!\cos\!\phi')\hat{\bfphi} \!+\!z'\hat{\textbf{z}}\big]}{\big( \rho'^2 + \rho^2 - 2\rho'(x\cos\phi' +y\sin\phi') +z'^2\big)^{3/2}}.
\end{align}
Outside the solenoid the potential is $\textbf{A}(\textbf{x}')=\Phi\hat{\bfphi}/(2\pi \rho'),$ which is used together with the unit vector outward to the surface $\hat{\textbf{n}}=\hat{\bfrho}$ and Eq.~(C2) to give the results
\begin{align}
\hat{\textbf{n}}\big(\textbf{E}\!\cdot\!\textbf{A}\big) =& \frac{q \,\Phi\,(x\sin\phi'-y\cos\phi')\hat{\bfrho}}{2 \pi \rho'\big( \rho^2\!+\!\rho'^2 \!-\! 2\rho'(x\cos\phi'\!+\!y\sin\phi')\!+\!z'^2\big)^{3/2}},\\
\textbf{A}(\hat{\textbf{n}}\!\cdot\!\textbf{E})=& \frac{q\,\Phi \big( \rho' - (x\cos\phi' +y\sin\phi') \big)\hat{\bfphi}}{2 \pi \rho'\big( \rho^2\!+\!\rho'^2 \!-\! 2\rho'(x\cos\phi'\!+\!y\sin\phi')\!+\!z'^2\big)^{3/2}},\\
\textbf{E}(\hat{\textbf{n}}\!\cdot\!\textbf{A})=&\, 0.
\end{align}
Using the position vector $\textbf{x}=\rho'\,\hat{\bfrho}$ and inserting Eqs.~(C3), C(4) and (C5) we obtain
\begin{align}
\nonumber \textbf{x}' \times &\big[ \hat{\textbf{n}}(\textbf{E}\!\cdot\!\textbf{A})-\textbf{A}(\hat{\textbf{n}} \cdot \textbf{E})-\textbf{E}(\hat{\textbf{n}}\!\cdot\!\textbf{A})\big]\\
=&-\frac{q\,\Phi \big( \rho' - (x\cos\phi' +y\sin\phi') \big)\hat{\textbf{z}}}{2 \pi \,\big( \rho^2+\rho'^2 - 2\rho'(x\cos\phi'+y\sin\phi')+z'^2\big)^{3/2}}\,.
\end{align}
Inserting Eq.~(C6) in Eq.~(C1) and using $dS'=\rho'\,d\phi'\,dz'$ it follows
\begin{align}
\nonumber \textbf{L}^\texttt{S}_{\texttt{AB}}=&  \lim_{\rho'\rightarrow \infty}\bigg(\!\!-\frac{q\Phi \rho'\,\hat{\textbf{z}}}{8\pi^2 c}\bigg)\\
&\times\int \frac{\big( \rho' - (x\cos\phi' +y\sin\phi') \big)\,d\phi'\,dz'}{\big( \rho^2\!+\!\rho'^2 \!-\! 2\rho'(x\cos\phi'\!+\!y\sin\phi')\!+\!z'^2\big)^{3/2}}\,.
\end{align}
The $z'$ integral is straightforward to evaluate via a change of variable. We obtain
\begin{align}
\nonumber\int^{+\infty}_{-\infty} &\frac{\big( \rho' - (x\cos\phi' +y\sin\phi') \big)\,dz'}{\big( \rho^2+\rho'^2 - 2\rho'(x\cos\phi'+y\sin\phi')+z'^2\big)^{3/2}}\\=&\frac{2\big( \rho' - (x\cos\phi' +y\sin\phi') \big)}{\big( \rho^2+\rho'^2 - 2\rho'(x\cos\phi'+y\sin\phi')\big)}\,.
\end{align}
Therefore
\begin{align}
\nonumber \textbf{L}^\texttt{S}_{\texttt{AB}}=&\lim_{\rho'\rightarrow \infty}\bigg(\!\!-\frac{q\Phi \rho'\,\hat{\textbf{z}}}{4\pi^2 c}\bigg)\\
&\times\oint\frac{\big( \rho' - (x\cos\phi' +y\sin\phi') \big)\,d\phi'}{\big( \rho^2+\rho'^2 - 2\rho'(x\cos\phi'+y\sin\phi')\big)}\,.
\end{align}
The $\phi'$ integral can be evaluated via a contour integration in the complex plane. We obtain the result
\begin{align}
\oint\frac{\big( \rho' - (x\cos\phi' +y\sin\phi') \big)\,d\phi'}{\big( \rho^2+\rho'^2 - 2\rho'(x\cos\phi'+y\sin\phi')\big)}=\frac{2\pi n}{\rho'}\, ,
\end{align}
where $n$ is the winding number of the charge path and we have assumed $\rho'^2>\rho^2>0$. Inserting Eq.~(C10) in Eq.~(C9) and after taking the limit $\rho'\rightarrow\infty$ we obtain
\begin{align}
\textbf{L}^\texttt{S}_{\texttt{AB}}=- \frac{nq\Phi\hat{\textbf{z}}}{2\pi c} \lim_{\rho'\rightarrow \infty}\bigg(\frac{\rho'}{\rho'}\bigg) = - \frac{nq\Phi\hat{\textbf{z}}}{2\pi c}.
\end{align}
Therefore $\textbf{L}^\texttt{S}_{\texttt{AB}}= - \textbf{L}_{\texttt{AB}}$ as expected.

{}

\end{document}